\documentclass[twocolumn,aps,floatfix,showpacs]{revtex4}
\usepackage[dvips]{graphicx}
\begin{document}
\title{Numerical study of the isotope effect in underdoped high-temperature
superconductors: Calculation of the angle-resolved photoemission spectra.}
\author{
A.S.~Mishchenko$^{1,2}$ and N.~Nagaosa$^{1,3}$}
\affiliation{$^1$CREST, Japan Science and Technology Agency (JST), 
AIST, 1-1-1, Higashi, Tsukuba 305-8562, Japan 
\\ 
$^2$RRC ``Kurchatov Institute", 123182, Moscow, Russia 
\\
$^3$CREST, Department of Applied Physics, The University of Tokyo, 7-3-1
Hongo, Bunkyo-ku, Tokyo 113, Japan}

\begin{abstract}
We present a numerical study of the isotope effect on the angle resolved 
photoemission spectra (ARPES) in the undoped cuprates. 
By the systematic-error-free Diagrammatic Monte Carlo method,
the Lehman spectral function of a single hole in the $tt't''-J$ model 
in the regime of intermediate and strong couplings to optical phonons 
is calculated for normal and isotope substituted systems.  
We found that the isotope effect is strongly energy-momentum dependent, 
and is anomalously enhanced in the intermediate coupling regime 
while it approaches 
to that of the localized hole model in the strong coupling regime. 
We predict the strengths of effect as well as the fine details of the 
ARPES lineshape change.  Implications to the doped case are also discussed. 
\end{abstract}


\maketitle

It is a subject of considerable debates for many years whether the
electron-phonon interaction (EPI) is essential for the physics of undoped 
and heavily underdoped high $T_c$ superconductors
\cite{Anderson,Marsh_96,Nor_98,Shen_03,Kyle_Dop,KyleSci_05}. 
Early study of the small isotope effect (IE) on $T_c$ at optimal doping 
together with the no prominent phonon features in the temperature dependence
of the resistivity $\rho(T)$ had led to the conclusion that the phonons are a
by-player in high $T_c$ cuprates. 
However, the roles of EPI have been recently studied intensively in terms 
of the neutron scattering, Raman scattering, the IE on $T_c$ and the 
superfluid density $\rho_s$, and ARPES \cite{Shen_03}. 
Especially ARPES provides the direct information on the single-particle 
Green function, which has revealed the ``kink'' structure of the electron 
energy dispersion 40-70 meV below the Fermi energy. 
The origin of this kink is naturally interpreted as the coupling to
some bosonic modes. 
The magnetic resonance mode and the phonon modes are the two major candidates, 
and the IE on ARPES should be the smoking-gun experiment to distinguish 
between these two. 
Gweon et al. \cite{Gweon_04} performed the ARPES experiment on 
O$^{18}$-replaced Bi2212 at optimal doping and found an appreciable IE, 
which however can not be explained within the conventional weak-coupling 
Migdal-Eliashberg theory. 
Namely the change of the spectral function due to O$^{18}$-replacement has 
been observed at higher energy region beyond the phonon energy ($\sim 60$meV). 
This is in sharp contrast to the weak coupling theory prediction, i.e., 
the IE should occur only near the phonon energy. 
Hence the IE in optimal Bi2212 remains still a puzzle. 
On the other hand, the ARPES in undoped materials, i.e., the single hole 
Lehman spectral function (LSF) doped into the Mott insulator, 
\cite{ZX95} has recently been understood in terms of the small polaron
formation  \cite{tJph,RoGuZX,RoGu2005}.
It has been revealed that the energy dispersion of the extended $t-J$ model,
$tt't''-J$ model \cite{Xiang_96} survives as the center of mass 
position of the broad Franck-Condon peak (FCP) of phonon side-bands 
even at strong EPI, while the weight $Z$ of the zero-phonon line and 
its dispersion is very small.
Recently, decoupling of chemical potential from observed in ARPES resonance, 
predicted by strong EPI scenario, was confirmed experimentally \cite{Kyle_Dop} 
and, thus, phonon origin of ARPES broadening seems to be likely. 
 
In addition to high-$T_c$ problem, strong EPI mechanism of ARPES
spectra broadening was considered as one of alternative scenarios
for diatomic molecules \cite{Zawatzky_89},
colossal magnetoresistive manganites \cite{Dessau_98},
quasi-one-dimensional Peierls conductors \cite{Perf_01},
and Verwey magnetites \cite{Schrupp_05}.
Therefore, exact analysis of the IE on ARPES at strong EPI is of
general interest for conclusive experiments in a broad variety 
of compound classes.     
 
In this Letter we present a study of the IE on the single particle
LSF of a hole strongly interacting with phonons in the $tt't''-J$ model, 
which is equivalent to study of ARPES in undoped cuprates \cite{Shen_03}.
As mentioned above, it is essential to compare experiment in undoped 
systems with the present DMC results, where theory can offer 
quantitative approximation-free results. 
We calculate IE on LSF by exact DMC method at zero temperature
in nodal $(\pi/2,\pi/2)$ and 
antinodal $(\pi,0)$ points of the Brillouin zone for realistic 
parameters of the $tt't''-J$ model in the intermediate and strong 
EPI regime.
 We find that IE is anomalously enhanced in the intermediate EPI regime 
but approaches simple analytic estimates in the strong EPI regime.  
We analyze the features of the FCP which are most 
sensitive to IE at different EPI regimes. 

In the standard spin-wave approximation for two-dimensional 
$tt't''-J$ model, which was shown to be a good approximation 
for small exchange integrals $J/t \le 0.4$ \cite{kane}, 
a hole with dispersion \cite{Xiang_96}
$\varepsilon ({\bf k}) = 4t'\cos(k_x)\cos(k_y) +
2t''(\cos(2k_x)+\cos(2k_y))$
propagates in the magnon and phonon (annihilation operators are 
$\alpha_{\bf k}$ and $b_{\bf k}$, respectively) bathes
\begin{equation}
\hat{H}_{\mbox{\scriptsize t-J}}^{0} =
\sum_{\bf k} \varepsilon ({\bf k}) h_{\bf k}^{\dagger} h_{\bf k} 
+
\sum_{\bf k} \omega_{\bf k} \alpha_{\bf k}^{\dagger} \alpha_{\bf k}
+
\Omega \sum_{\bf k} b_{\bf k}^{\dagger} b_{\bf k}
\label{h0}
\end{equation}
with magnon dispersion $\omega_{\bf k}=2J\sqrt{1-\gamma_{\bf k}^2}$, where
$\gamma_{\bf k}=(\cos k_x + \cos k_y) / 2$. The hole is scattered by magnons 
\begin{equation}
\hat{H}_{\mbox{\scriptsize t-J}}^{\mbox{\scriptsize h-m}} =
N^{-1/2} \sum_{\bf k , q} M_{\bf k , q} 
\left[ h_{\bf k}^{\dagger} h_{\bf k-q} \alpha_{\bf q} + h.c.
\right] \; ,
\label{h-m}
\end{equation}  
where $M_{\bf k , q}$ is the standard vertex \cite{kane}. 
We chose the simplest Holstein short-range EPI Hamiltonian  
\begin{equation}
\hat{H}^{\mbox{\scriptsize e-ph}} = 
N^{-1/2}  \sum_{\bf k , q} \frac{\sigma}{\sqrt{2M\Omega}} 
\left[ h_{\bf k}^{\dagger} h_{\bf k-q} b_{\bf q} + h.c.
\right] \; ,
\label{e-ph}
\end{equation}
where $\sigma$ is the momentum and isotope independent coupling 
constant, 
$M$ is the mass of the vibrating lattice ions, and 
$\Omega$ is the frequency of dispersionless phonon 
(Planck constant $\hbar$ is set to unity). 
We introduce dimensionless coupling constant 
$\lambda=\gamma^2/4t\Omega$ 
which is, in contrast to the standard Holstein constant 
$\gamma=\sigma/\sqrt(2M\Omega)$, is invariable quantity for the 
simplest case of IE. 
Indeed, assuming the simplest natural relation $\Omega \sim 1/\sqrt{M}$ 
between phonon frequency and mass, we find that $\lambda$ does not depend 
on the isotope factor 
$\kappa_{\mbox{\scriptsize iso}} = \Omega/\Omega_0 = \sqrt{M_0/M}$, which 
is defined as the ratio of phonon frequency in isotope substituted 
$\Omega$ and normal $\Omega_0$ system.
\begin{figure}[hbt]
\hspace{-0.45 cm}  \vspace {-0.5 cm}
\includegraphics{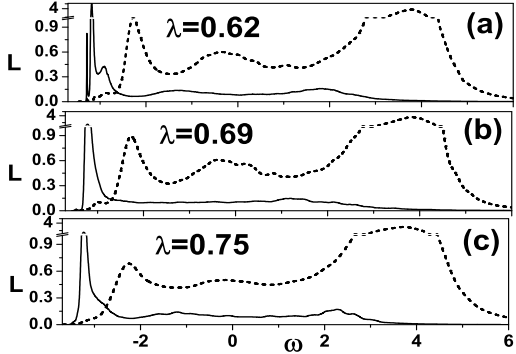}
\caption{\label{fig:fig1} Hole LSFs at ${\bf k}=(\pi/2,\pi/2)$ (solid line)
and ${\bf k}=(\pi,0)$ (dashed line) for different couplings.  
} 
\end{figure}
\begin{figure}[htb]
\hspace{0.8 cm}  \vspace {-0.5 cm}
\includegraphics{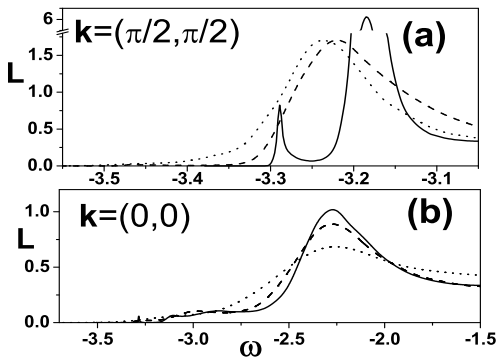}
\caption{\label{fig:fig2} Low energy part of LSF for $\lambda=0.62$
(solid line), $\lambda=0.69$ (dashed line), and $\lambda=0.75$ 
(dotted line) at nodal (a) and antinodal (b) points.}
\end{figure}

We chose adopted parameters of the $tt't''-J$ model which reproduce 
the experimental dispersion of a quasiparticle (QP) \cite{Xiang_96}: 
$J/t=0.4$, $t'/t=-0.34$, and $t''/t=0.23$ . 
The frequency of the relevant phonon \cite{Shen_03} is set to 
$\Omega/t=0.2$ and the isotope factor 
$\kappa_{\mbox{\scriptsize iso}}=\sqrt{16/18}$
corresponds to substitution of O$^{18}$ isotope for O$^{16}$. 
  
We use DMC method \cite{MPSS,MN01_Opt03}, which is the only available 
method for approximation-free study of excited states of problem 
(\ref{h0}-\ref{e-ph}) for the macroscopic system at strong 
EPI. 
Non-crossing approximation for phonon variables \cite{Ramsak_92} 
is shown to be invalid approximation for strong EPI \cite{tJph}  while 
exact diagonalization studies of small clusters, although account for 
correct treatment of phonons, imply a discrete spectrum and, thus, 
tiny changes of the FCP lineshape for $\kappa \approx 1$ are not 
reliable. 
The only study of IE by exact method \cite{KorAl_05}, i.e. path integral 
quantum Monte Carlo algorithm \cite{SpensKor_05}, does not addresses realistic 
$tt't''-J$ model, where both phonon and magnon variables have to be 
summed simultaneously and does not study ARPES directly. 
On the other hand, DMC method allows to treat ARPES directly \cite{tJph}. 
We use recently developed Stochastic Optimization method \cite{MPSS} 
which avoids regularization and artificial broadening of LSF peaks.
To sweep aside any doubts of possible instabilities of analytic 
continuation, we calculate the LSF for normal compound 
($\kappa_{\mbox{\scriptsize nor}}=1$), isotope substituted 
($\kappa_{\mbox{\scriptsize iso}}=\sqrt{16/18}$) 
and ``anti-isotope'' substituted 
($\kappa_{\mbox{\scriptsize ant}}=\sqrt{18/16}$) compounds. 
Monotonic dependence of LSF on $\kappa$ ensures stability of analytic 
continuation and gives possibility to evaluate the error-bars of a quantity
${\cal A}$ using quantities 
${\cal A}_{\mbox{\scriptsize iso}}-{\cal A}_{\mbox{\scriptsize nor}}$,       
${\cal A}_{\mbox{\scriptsize nor}}-{\cal A}_{\mbox{\scriptsize ant}}$, and 
$({\cal A}_{\mbox{\scriptsize iso}}-{\cal A}_{\mbox{\scriptsize ant}})/2$.

Figure \ref{fig:fig1} presents LSFs in nodal and antinodal points for values 
of EPI which are larger than the critical self-trapping (ST) 
coupling $\lambda_{\mbox{\scriptsize cr}}\approx 0.58$, i.e. coupling 
where fast transformation to regime of strong EPI occurs
\cite{Rashba82,RP}. 
Since LSF is sensitive to strengths of EPI only for low frequencies 
we concentrate on the low energy part of the spectrum 
(Fig.~\ref{fig:fig2}). 
It is seen that near the ST crossover the LSF in 
the nodal point quickly changes with coupling while FCP in antinodal point 
gradually broadens. The bandwidth of FCP dispersion is estimated as 
$W_{\mbox{\scriptsize FCP}} \approx 0.9$. 

\begin{figure}[htb]
\hspace{-18 mm}  \vspace {-0.5 cm}
\includegraphics{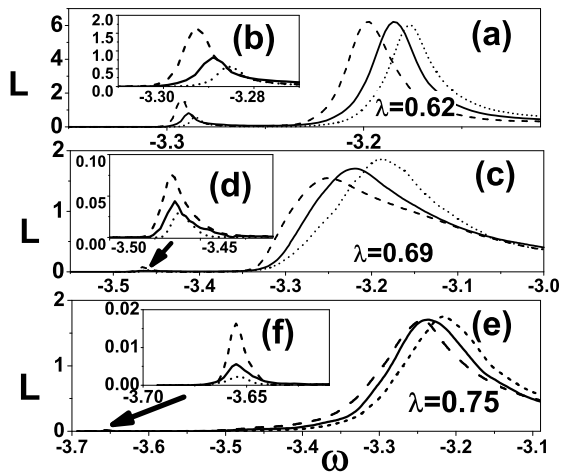}
\caption{\label{fig:fig3}  Low energy part of hole LSFs in the nodal 
point at different couplings (a, c, e): normal compound (solid line), 
isotope substituted compound (dotted line) and
``antiisotope'' substituted compound (dashed line).
Insets (b, d, e) show low energy real QP peak. }
\end{figure}

\begin{figure}[hbt]
\hspace{-20 mm}  \vspace {-0.5 cm}
\includegraphics{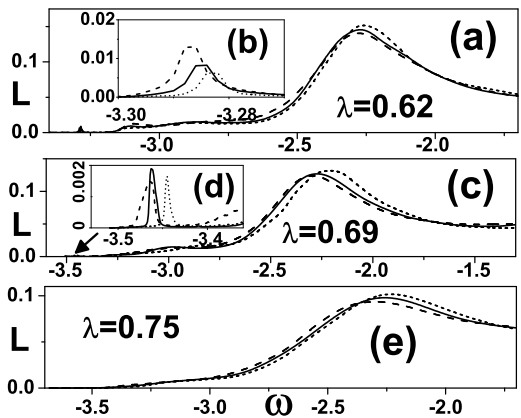}
\caption{\label{fig:fig4}  Low energy part of hole LSFs in the antinodal 
point. See caption of Fig.~\ref{fig:fig3}. }
\end{figure}

Figures \ref{fig:fig3} and \ref{fig:fig4} show IE on the hole LSF 
for different couplings in nodal and antinodal points, respectively. 
The general trend is a shift of all spectral features to larger 
energies with increase of the isotope mass ($\kappa<1$). 
One can also note that the shift of broad FCP is much larger than that 
of real narrow QP peak. 
Moreover, for large couplings $\lambda$ the shift of QP energy 
approaches zero and only decrease of QP spectral weight $Z$ 
is observed for larger isotope mass. 
On the other hand, the shift of FCP is not suppressed for larger 
couplings. Except for the LSF in nodal point at $\lambda=0.62$ 
(Fig.~\ref{fig:fig3}a, b), where LSF still has a notable weight of QP 
$\delta$-functional peak, there is one more notable feature of the IE.
With increase of the isotope mass the height of FCP increases.
Taking into account the conservation law for LSF 
$\int_{-\infty}^{+\infty}L_{\bf k}(\omega)=1$ and insensitivity 
of high energy part of LSF to EPI strength (Fig.~\ref{fig:fig2}), 
the narrowing of the FCP for larger isotope mass can be concluded.
To understand the trends of the IE in the strong coupling regime
we analyze the exactly solvable independent oscillators model
(IOM) \cite{Mahan}.
More rigorous Lang-Firsov transformation is not required since 
it is not valid for adiabatic case $\Omega/t=0.2$ and in the intermediate 
coupling regime \cite{KorAl_05} while in the strong coupling regime, 
as it is shown below, IOM shows quantitative agreement with DMC data. 
Indeed, coherent QP bandwidths 
$W_{\mbox{\scriptsize QP}}(\lambda=0.62) \approx 3.2*10^{-3}$,  
$W_{\mbox{\scriptsize QP}}(\lambda=0.69) \approx 6*10^{-4}$, and 
$W_{\mbox{\scriptsize QP}}(\lambda=0.75) \approx 4*10^{-4}$
are negligibly small in comparison with FCP bandwidth 
$W_{\mbox{\scriptsize FCP}} \approx 0.9$.
\begin{figure}[tbh]
\hspace{-0.3 cm}  \vspace {-0.5 cm}
\includegraphics{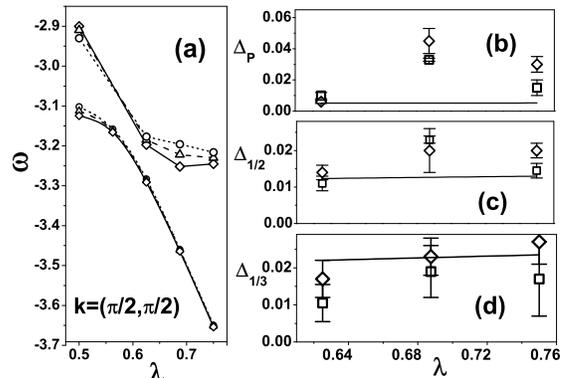}
\caption{\label{fig:fig5} 
 Energies of ground state and broad peaks
(a) for normal (tringles), isotope substituted
(circles) and ``antiisotope'' substituted (dimonds) compounds.
Comparison of IOM estimates (lines) with 
DMC data in the nodal (squares) and antinodal (diamonds) points:
shift of the FCP top (b), FCP leading edge at $1/2$ of height (c),
and FCP leading edge at $1/3$ of height (d).}
\end{figure}
The LSF in IOM is the Poisson distribution
\begin{equation}
L(\omega) = \exp[-\xi_0/\kappa] \sum_{l=0}^{\infty} 
\frac{[\xi_0/\kappa]^l}{l!} {\cal G}_{\kappa,l}(\omega) \; ,
\label{lsf}
\end{equation}
where $\xi_0=\gamma^2_0/\Omega_0^2=4t\lambda/\Omega_0$ is 
dimensionless coupling constant for normal system and 
${\cal G}_{\kappa,l}(\omega) = \delta[\omega+4t\lambda-\Omega_0\kappa l]$
is the $\delta$-function. 
The properties of the Poisson distribution quantitatively explain 
many features of the IE on LSF. 

The energy $\omega_{\mbox{\scriptsize QP}}=-4t\lambda$ of the zero-phonon 
line $l=0$ in (\ref{lsf}) depends only on isotope independent quantities
which explains very weak isotope dependence of QP peak energy in insets of  
Figs.~\ref{fig:fig3}-\ref{fig:fig4}. 
Besides, change of the zero-phonon line weight 
$Z^{(0)}$ obeys relation 
$Z^{(0)}_{\mbox{\scriptsize iso}}/Z^{(0)}_{\mbox{\scriptsize nor}} =
\exp\left[ -\xi_0(1-\kappa)/\kappa\right]$ in IOM. 
These IOM estimates agree with DMC data within 15\% in the nodal point and 
within 25\% in the antinodal one.  
IE on FCP in the strong coupling regime follows from the properties 
of zero
$M_0=\int_{-\infty}^{+\infty} L(\omega) d \! \omega =1$,
first 
$M_1=\int_{-\infty}^{+\infty} \omega L(\omega) d \! \omega =0$, and second
$M_2=\int_{-\infty}^{+\infty} \omega^2 L(\omega) d \! \omega 
= \kappa \xi_0 \Omega_0^2$ moments of shifted Poisson distribution 
(\ref{lsf}). 
Moments $M_0$ and $M_2$ establish relation  
${\cal D} = h^{\mbox{\scriptsize FCP}}_{\mbox{\scriptsize iso}} /
h^{\mbox{\scriptsize FCP}}_{\mbox{\scriptsize nor}} = 1/\sqrt{\kappa}
\approx 1.03$ between heights of FCP in normal and substituted compounds.
DMC data in the antinodal point perfectly agree with the above estimate 
for all couplings. 
This is consistent with the idea that the anti-nodal region remains 
in the strong coupling regime even though the nodal region is in the 
crossover region.
In the nodal point DMC data well agree with IOM estimate for 
$\lambda=0.75$ 
(${\cal D} \approx 1.025$) whereas at $\lambda=0.69$ and $\lambda=0.62$ 
influence of the ST point leads to anomalous values of 
${\cal D}$: ${\cal D} \approx 1.07$ and ${\cal D} \approx 0.98$, respectively.
Shift of the low energy edge at half maximum $\Delta_{1/2}$ must be  
proportional to change of the root square of second moment 
$\Delta_{\sqrt{M_2}} = \sqrt{\xi_0} \Omega_0 [1-\sqrt{\kappa}]$. 
As we found in numeric simulations of Eq.~(\ref{lsf}) with Gaussian 
functions \cite{Gauss} ${\cal G}_{\kappa,l}(\omega)$, 
relation $\Delta_{1/2} \approx \Delta_{\sqrt{M_2}} / 2$ is accurate to 10\% 
for $0.62<\lambda<0.75$. 
Also, simulations show that the shift of the edge at one third of 
maximum $\Delta_{1/3}$ obeys relation 
$\Delta_{1/3} \approx \Delta_{\sqrt{M_2}}$. 
DMC data with IOM estimates are in good agreement for strong EPI 
$\lambda=0.75$ (Fig.~\ref{fig:fig5}). However, shift of the FCP top 
$\Delta_p$ and $\Delta_{1/2}$ are considerably enhanced in the 
self-trapping (ST) transition region. 
The physical reason for enhancement of IE in this region
is general property regardless of the QP dispersion, range of EPI, etc.  
The influence of nonadiabatic matrix element, mixing excited and ground 
states, on the energies of resonances essentially depends on the 
phonon frequency. 
While in the adiabatic approximation ST transition is sudden and 
nonanalytic in $\lambda$ \cite{Rashba82}, nonadiabatic matrix elements turn 
it to smooth crossover \cite{Gerlach}. 
Thus, as illustrated in Fig.~\ref{fig:fig5}a, the smaller the frequency 
the sharper the kink in the dependence of excited state energy on the 
interaction constant.

Cautions should be made about approximate form of EPI (\ref{e-ph}). 
Strictly speaking, actual momentum dependence of the interaction
constant $\sigma$ \cite{RoGu_04,Ishihara_04} can slightly change the
obtained differences between nodal and antinodal points though
the general trends have to be left intact because ST is caused 
solely by the short range part of EPI \cite{Rashba82}. 
Also the shell model calculation \cite{RoGu_04}
has shown that the dominant EPI is the short range one associated with 
the oxygen displacements.
Therefore, even though the participation of Cu ions \cite{RoGu_04} in the 
lattice vibrations relevant to EPI violates simple relation 
$\Omega \sim 1/\sqrt{M}$, this relation is approximately valid.

Finally, we discuss the relevance of the present results to the isotope 
experiment on the doped high $T_c$ cuprates \cite{Gweon_04}. 
The most puzzling feature is that the IE is negligible near the phonon 
frequency, and is mainly observed in the high energy region. 
This can not be explained by the 
conventional Eliashberg-Migdal theory, which predicts the suppression of the 
multiphonon emission at high energy and the IE only near 
the one-phonon frequency.
The intermediate and strong coupling polaron theory gives a 
completely different picture, where the high energy part consists 
of the multiphonon sidebands and is 
subject to the IE as shown in this paper. 
Although it is not clear at the moment when the 
crossover from the doped hole picture to the metallic large Fermi surface 
occurs as the doping proceeds, the IE 
experiment \cite{Gweon_04} suggests the former 
might persists even at the optimal doping. However more study is needed both 
theoretically and experimentally for the doped case. In the undoped case,
on the other hand, the present results can be directly compared with the 
experiments and more solid studies can be done.
It is found that isotope effect on the ARPES lineshape of a single hole
is anomalously enhanced in the intermediate coupling regime while can be 
described by simple independent oscillators model in the strong 
coupling regime. 
The shift of FCP top and change of the FCP height are relevant quantities 
to pursue experimentally in the intermediate coupling regime since IE 
on these characteristics is enhanced near the self trapping point. 
In contrast, shift of the leading edge is the relevant quantity in the 
strong coupling regime since this value increases with coupling as 
$\sqrt{\lambda}$. 
These conclusions, depending on the fact whether self trapping phenomenon 
is encountered in specific case, can be applied fully or partially to another 
compounds with strong EPI 
\cite{Dessau_98,Perf_01,Schrupp_05}.

{\it Note added in proof.} Related studies of the isotope effect on  
ARPES were recently reported by S.\ Fratini and S.\ Ciuchi \cite{Fratini}
in the framework of dynamical mean field approach to the Mott-Hubbard 
insulator and by G.\ Seibold and M.\ Grilli \cite{Grilli} where 
coupling of quasiparticle to a charge collective mode is 
treated within the simple perturbative scheme. 

Fruitful discussions with G.-H.\ Gweon are acknowledged. 
This work was supported by RFBR 04-02-17363a, and 
the Grants-in-aid for Scientific Research and NAREGI Nanoscience Project 
from the Ministry of Education, Culture, Sports, Science, and Technology.


\begin{thebibliography}{99}
%
%
\bibitem{Anderson} P.\ W.\ Andreson, {\it The Theory of Superconductivity 
                   in the High-$T_c$ Cuprates} (Princeton University Press,
                   New Jersey, 1997). 
%
\bibitem{Marsh_96} D.\ S.\ Marshall, D. S. Dessau, A. G. Loeser, 
                  C-H. Park, A. Y. Matsuura, J. N. Eckstein, 
                  I. Bozovic, P. Fournier, A. Kapitulnik, 
                  W. E. Spicer, and Z.-X. Shen,
                   Phys.\ Rev.\ Lett.\ {\bf 76}, 4841 (1996).
%
\bibitem{Nor_98} M.\ R.\ Norman, H. Ding, M. Randeria, J.C. Campuzano, 
                 T. Yokoya, T. Takeuchi, T. Takahashi, T. Mochiku,  
                   K. Kadowaki, P. Guptasarma, and D.G Hinks,
                Nature {\bf 392}, 157 (1998) 
%
\bibitem{Shen_03} A.\ Danmascelli, Z.-X.\ Shen, and Z.\ Hussain,
         Rev.\ Mod.\ Phys.\ {\bf 75}, 473 (2003).
%
\bibitem{Kyle_Dop} K.\ M.\ Shen, F. Ronning, D. H. Lu, W. S. Lee,
                    N. J. C. Ingle, W. Meevasana, F. Baumberger,
                    A. Damascelli, N. P. Armitage, L. L. Miller, 
                     Y. Kohsaka, M. Azuma, M. Takano, H. Takagi, 
                     and Z.-X. Shen,  
                   Phys.\ Rev.\ Lett.\ {\bf 93}, 267002 (2004).
%
\bibitem{KyleSci_05} K.\ M.\ Shen, F. Ronning, D. H. Lu,
                     F. Baumberger, N. J. C. Ingle, W. S. Lee, 
                     W. Meevasana, Y. Kohsaka, M. Azuma, M. Takano, 
                     H. Takagi, and Z.-X. Shen,  
                     Science {\bf 307}, 901 (2005).
%
\bibitem{Gweon_04} G.-H.\ Gweon, T. Sasagawa, S. Y. Zhou, J. Graf, H. Takagi, 
                   D. H. Lee, and A. Lanzara,  
                   Nature {\bf 430}, 187 (2004).
%
\bibitem{ZX95} B.\ O.\ Wells, Z.-X. Shen, A. Matsuura, D. M. King, 
               M. A. Kastner, M. Greven, and R. J. Birgeneau,
         Phys.\ Rev.\ Lett.\ {\bf 74}, 964 (1995).
%
\bibitem{tJph} A.\ S.\ Mishchenko and N.\ Nagaosa, 
               Phys.\ Rev.\ Lett.{\bf 93}, 036402 (2004).
%
\bibitem{RoGuZX} O.\ R\"{o}sch, O. Gunnarsson, X. J. Zhou, T. Yoshida, 
                T. Sasagawa, A. Fujimori, Z. Hussain, Z.-X. Shen, 
               and S. Uchida, 
               Phys.\ Rev.\ Lett.\ {\bf 95}, 227002 (2005).
%
\bibitem{RoGu2005} O.\ R\"{o}sch and O.\ Gunnarsson,
                   Eur.\ Phys.\ J B {\bf 43}, 11 (2005).
%
\bibitem{Xiang_96} T.\ Xiang and M.\ Wheatley, Phys.\ Rev.\ B {\bf 54},
         R12653 (1996).
%
\bibitem{Zawatzky_89} G.\ A.\ Sawatzky, 
                      Nature (London) {\bf 342B}, 480 (1989).
%
\bibitem{Dessau_98} D.\ S. Dessau, T. Saitoh1, C.-H. Park, Z.-X. Shen, 
                    P. Villella, N. Hamada, Y. Moritomo, and Y. Tokura,
                   Phys.\ Rev.\ Lett.\ {\bf 81}, 192 (1998);
                   V.\ Perebeinos and P.\ B.\ Allen, 
                   {\it ibid}.\ {\bf 85}, 5178 (2000).
                   N.\ Mannella, A. Rosenhahn, C. H. Booth, S. Marchesini,
                   B. S. Mun, S.-H. Yang, K. Ibrahim, Y. Tomioka, and 
                   C. S. Fadley,  
                   {\it ibid}.\ {\bf 92}, 166401 (2004). 
%
\bibitem{Perf_01} L.\ Perfetti,H. Berger, A. Reginelli, L. Degiorgi, 
                   H. Höchst, J. Voit, G. Margaritondo, and M. Grioni, 
                  Phys.\ Rev.\ Lett.\ {\bf 87}, 216404 (2001);
                  L. Perfetti, S. Mitrovic, G. Margaritondo, M. Grioni,
                  L. Forr\'{o}, L. Degiorgi, and  H. H\"{o}chst, 
                  Phys.\ Rev.\ B {\bf 66}, 075107 (2002).
%
\bibitem{Schrupp_05} D.\ Schrupp, M. Sing, M. Tsunekawa, H. Fujiwara, 
                     S. Kasai, A. Sekiyama, S. Suga, T. Muro, 
                     V. A. M. Brabers and R. Claessen,  
                     Eur.\ Phys.\ Lett.\ {\bf 70}, 789 (2005).
%
\bibitem{kane} C.\ L. Kane, P.\ A.\ Lee, and N.\ Read, 
         Phys.\ Rev.\ B {\bf 39}, 6880 (1989); 
         Z.\ Liu and E.\ Manousakis, 
         {\it ibid}.\ {\bf 45}, 2425 (1992).
%
\bibitem{MPSS} A.\ S.\ Mishchenko, N. V. Prokof'ev, A. Sakamoto, 
               and B. V. Svistunov, 
               Phys.\ Rev.\ B {\bf 62}, 6317 (2000).
%
\bibitem{MN01_Opt03} A.\ S.\ Mishchenko and N.\ Nagaosa, 
               Phys.\ Rev.\ Lett. {\bf 86}, 4624 (2001);
               A.\ S.\ Mishchenko, N. Nagaosa, N. V. Prokof'ev, 
               A. Sakamoto, and B. V. Svistunov,
               {\it ibid}.\ {\bf 91}, 236401 (2003); 
               A.\ S.\ Mishchenko, 
               Physics-Uspekhi {\bf 48}, 887 (2005).
%
\bibitem{Ramsak_92} A.\ Ramsak, P.\ Horsch, and P.\ Fulde, 
                    Phys.\ Rev.\ B {\bf 46}, 14305 (1992);
                    B.\ Kyung, S. I. Mukhin, V. N. Kostur, and 
                    R. A. Ferrell,  
                    {\it ibid}.\ {\bf 54}, 13167 (1996).
%
\bibitem{KorAl_05} P.\ E.\ Kornilovitch and A.\ S.\ Alexandrov,
                   Phys.\ Rev.\ B {\bf 70}, 224511 (2004).
%
\bibitem{SpensKor_05} P.\ E.\ Spenser,  J. H. Samson, P. E. Kornilovitch,
                    and A. S. Alexandrov, 
                   Phys.\ Rev.\ B {\bf 71}, 184310 (2005). 
%
\bibitem{Rashba82} E.\ I.\ Rashba, in
         {\it Modern Problems in Condensed Matter Sciences}, Ed.\ by
         V.\ M.\ Agranovich and A.\ A.\ Maradudin (North-Holland, 
         Amsterdam, 1982), Vol.\ 2, p.\ 543. 
%
\bibitem{RP} A.S.\ Mishchenko, N. Nagaosa, N. V. Prokof'ev, A. Sakamoto, 
             and B. V. Svistunov, 
              Phys. Rev. B {\bf 66}, 020301 (2002).
%
\bibitem{Mahan} G.\ D.\ Mahan, {\it Many Particle Physics} (Plenum Press,
                New york, 1990).
%
\bibitem{Gerlach} B.\ Gerlach and H.\ L\"owen, 
               Rev.\ Mod.\ Phys.\ {\bf 63}, 63 (1991).
%
\bibitem{Gauss} Results are almost independent on the parameter
$\eta$ of the Gausssian distribution
${\cal G}_{\kappa,l}(\omega) 
= 1/(\eta\sqrt{2\pi}) \exp(-[\omega+4t\lambda-\Omega_0\kappa l]/(2\eta^2))$ 
in the range $[0.12,0.2]$.  
%
\bibitem{RoGu_04} O.\ R\"{o}sch and O.\ Gunnarsson,
                  Phys.\ Rev.\ Lett.\ {\bf 92}, 146403 (2004).
%
\bibitem{Ishihara_04} S.\ Ishihara and N.\ Nagaosa,
                      Phys.\ Rev.\ B {\bf 69}, 144520 (2004). 
%
\bibitem{Fratini} S.\ Fratini and S.\ Ciuchi, 
                  Phys.\ Rev.\ B {\bf 72}, 235107 (2005).
%
\bibitem{Grilli} G.\ Seibold and M.\ Grilli,
                  Phys.\ Rev.\ B {\bf 72}, 104519 (2005). 
\end{thebibliography}
\end{document}